\documentclass[a4paper,12pt]{article}
\def\bea{\begin{array}}
\def\bem{\begin{displaymath}}
\def\beq{\begin{equation}}

\def\eea{\end{array}}
\def\eem{\end{displaymath}}
\def\eeq{\end{equation}}

\def\ov{\overline}

\def\s2w{\sin^2 \theta_W}

\def\crbig{\\\noalign{\vspace {3mm}}}

\def\be{\begin{equation}}
\def\ee{\end{equation}}
\def\bc{\begin{center}}
\def\ec{\end{center}}
\def\bea{\begin{eqnarray}}
\def\eea{\end{eqnarray}}
\def\dd{\displaystyle}
\def\nn{\nonumber}

\catcode`@=11
\def\marginnote#1{}
\newcount\hour
\newcount\minute
\newtoks\amorpm
\hour=\time\divide\hour by60
\minute=\time{\multiply\hour by60 \global\advance\minute by-\hour}
\edef\standardtime{{\ifnum\hour<12 \global\amorpm={am}%
        \else\global\amorpm={pm}\advance\hour by-12 \fi
        \ifnum\hour=0 \hour=12 \fi
        \number\hour:\ifnum\minute<10 0\fi\number\minute\the\amorpm}}
\edef\militarytime{\number\hour:\ifnum\minute<10 0\fi\number\minute}
\def\draftlabel#1{{\@bsphack\if@filesw {\let\thepage\relax
   \xdef\@gtempa{\write\@auxout{\string
      \newlabel{#1}{{\@currentlabel}{\thepage}}}}}\@gtempa
   \if@nobreak \ifvmode\nobreak\fi\fi\fi\@esphack}
        \gdef\@eqnlabel{#1}}
\def\@eqnlabel{}
\def\@vacuum{}
\def\draftmarginnote#1{\marginpar{\raggedright\scriptsize\tt#1}}
\def\draft{\oddsidemargin 0.0truein
        \def\@oddfoot{\sl preliminary draft \hfil
        \rm\thepage\hfil\sl\today\quad\militarytime}
        \let\@evenfoot\@oddfoot \overfullrule 3pt
        \let\label=\draftlabel
        \let\marginnote=\draftmarginnote
   \def\@eqnnum{(\theequation)\rlap{\kern\marginparsep\tt\@eqnlabel}%
\global\let\@eqnlabel\@vacuum}  }
\catcode`@=12
\begin{document}
\begin{titlepage}
\vspace*{-1cm}
\begin{flushright}
NEIP--04--08 \\
LPTENS--0440\\
CPTH-RR053.0904\\
CERN--PH--TH/2004--228 \\
ROMA--1395/04
\end{flushright}
\vspace{0.5cm}
\bc
{\Large\bf Superpotentials in IIA compactifications
\\ \vspace*{2.0mm} with general fluxes}
\ec
\vskip.5cm
\bc
{\large \bf J.-P.~Derendinger~$^1$, C.~Kounnas~$^{2 \, \diamond, \, 3}$,
\\ \vskip.4cm
P.~M.~Petropoulos~$^{4 \, \dagger}$ and F.~Zwirner~$^{3, \, 5}$}
\vskip .5cm
{\small
$^1$ Physics Institute, Neuch\^atel University, \\
Breguet 1, CH--2000 Neuch\^atel, Switzerland
\vskip .2cm
$^2$ Laboratoire de Physique Th\'eorique,
Ecole Normale Sup\'erieure, \\
24 rue Lhomond, F--75231 Paris Cedex 05, France
\vskip .2cm
$^3$ CERN, Physics Department, Theory Division,
\\
CH--1211 Geneva 23, Switzerland
\vskip .2cm
$^4$ Centre de Physique Th\'eorique, Ecole Polytechnique,
\\
F--91128 Palaiseau, France
\vskip .2cm
$^5$ Dipartimento di Fisica, Universit\`a di Roma `La Sapienza', and
\\ INFN, Sezione di Roma, P.le A.Moro 2, I--00185 Rome, Italy
}
\ec
\vspace{0.3cm}
\begin{abstract}
\noindent We derive the effective $N=1$, $D=4$ supergravity for
the seven main moduli of type IIA orientifolds with D6 branes,
compactified on $T^6/(Z_2 \times Z_2)$ in the presence of general
fluxes. We illustrate and apply a general method that relates the
$N=1$ effective K\"ahler potential and superpotential to a
consistent truncation of gauged $N=4$ supergravity. We identify
the correspondence between various admissible fluxes, $N=4$
gaugings and $N=1$ superpotential terms. We construct explicit
examples with different features: in particular, new IIA no-scale
models and a model which admits a supersymmetric $AdS_4$ vacuum
with all seven main moduli stabilized.
\end{abstract}
\vspace*{0.3cm}\vfill
\hrule width 6.7cm
\vskip.1cm{\small \small \small
$^\diamond$\  Unit{\'e} mixte  du CNRS et de l'Ecole Normale
Sup{\'e}rieure, UMR 8549. \\
$^\dagger$\ Unit{\'e} mixte  du CNRS et de  l'Ecole Polytechnique,
UMR 7644.}
\end{titlepage}
\setcounter{footnote}{0}
\vskip2truecm
\setlength{\baselineskip}{.7cm}
\setlength{\parskip}{.2cm}
\newpage
\section{Introduction}
\label{secintro}

Compactifications of superstrings and M-theory~\footnote{For an
introduction, see e.g. \cite{polbook}.} may lead to four-dimensional
vacua with exact or spontaneously broken supersymmetries. The pattern
of residual and broken supersymmetries strongly depends on the set of
moduli fields predicted by the compactification geometry and on the
detailed dynamics of these moduli. Even for the phenomenologically
attractive compactifications with spontaneously broken $N = 1$ only,
information on the dynamics of moduli is provided by the much larger
symmetry of the underlying $D=10$ string theories, with sixteen or
thirty-two supercharges. Similarly, in the effective $D=4$ low-energy
supergravity theory, this information on moduli dynamics is encoded in
the underlying $N \ge 4$ supersymmetry. Thus, the K\"ahler potential
of the $N=1$ effective supergravity follows from the scalar
sigma-model induced by $N=4$ auxiliary field and gauge-fixing
equations. And the $N=1$ superpotential for the moduli and matter
fields is directly related to the $N=4$ supergravity \cite{ungn4,
Chams, DF} gauging \cite{N=4}, which in turn corresponds to a specific
flux structure of the underlying ten-dimensional string theory or
eleven-dimensional $M$--theory.

The generation of a scalar potential for the moduli fields is a
crucial ingredient in supersymmetry breaking and in the determination
of a stable $D=4$ background geometry, if any. It is also essential to
reduce the number of massless scalars and/or undetermined parameters
in the low-energy effective theory. Besides the curvature of the
internal space itself, there are several well-known sources for a
scalar potential in the compactified ten-dimensional (or
eleven-dimensional) theory.

A first source is the Scherk-Schwarz mechanism \cite{ftss}, and its
generalization to superstrings via freely acting orbifolds
\cite{strss}. The relevant fluxes are the geometrical ones, associated
with the internal spin connection $\omega_3$. Some of the
corresponding effective theories are no-scale supergravity models
\cite{noscale}, with broken supersymmetry in a flat $D=4$
background. However, the gravitino and the other masses generated in
this way are proportional (modulo quantized charges) to the inverse
length scale of the compactified space, $m \propto R^{-1}$. Therefore,
to have supersymmetry breaking and/or preserving TeV scale masses, we
need a very large internal dimension, $R \sim 10^{15} \, l_P$, where
$l_P$ is the (four-dimensional) Planck length.

A second source is non-zero ``fluxes" of antisymmetric tensor fields,
as first identified long ago for the three-form $H_3$ of the heterotic
theory \cite{h3het}. There is an extensive recent literature
\cite{IIBflu} on orientifolds of the IIB theory in the presence of
three-form fluxes. For instance, simultaneous and suitably
aligned NS-NS (NS $=$ Neveu-Schwarz) and R-R (R $=$ Ramond) 3-form
fluxes, $H_3$ and $F_3$, can lead to no-scale supergravities,
but now $m \propto l_P^{\, 2} \, R^{-3}$: as a result, TeV scale
supersymmetry breaking and/or preserving masses can be obtained for $R
\sim 10^{5} \, l_{P}$. The richer flux content of the IIA theory has
been studied to a lesser extent \cite{IIAflu, aft}.

Both sources, geometric and antisymmetric tensor fluxes, can be
combined, as originally examined in the heterotic theory by Kaloper
and Myers \cite{kalmy}.

In this paper, we use the method of supergravity gaugings to
describe in general terms the generation of moduli superpotentials
in a specific compactification scheme, defined as follows. We
consider compactifications of superstring theories on the orbifold
$T^6 / ( Z_2 \times Z_2 )$, combined for type-II strings with a
compatible orientifold projection to reduce supersymmetry to four
supercharges. The moduli spectrum includes then seven chiral
multiplets from the closed string sector, and the orbifold has a
natural permutation symmetry in the three two-tori ($T^2$) defined
by the action of $Z_2 \times Z_2$ on the six-torus $T^6$. We then
construct the gaugings associated to general flux structures
respecting this `plane-interchange' permutation symmetry (this
assumption could be eventually relaxed, leading to a wider
spectrum of possibilities). We include the fluxes generated by all
antisymmetric tensor fields (NS-NS and R-R), and also geometrical
fluxes associated to components of the internal spin connection,
as in Scherk-Schwarz compactifications. We analyze here in detail
the case of IIA strings (with $D6$--branes), since it offers the
broadest choice of fluxes and breaking patterns. We establish the
dictionary relating fluxes, gauging structure constants and
superpotential terms, and the consistency conditions applying on
gauging and flux coefficients. This general formulation allows us
to study examples with selected phenomenological properties.  We
find in particular that gaugings and fluxes exist in IIA
compactifications, such that all seven moduli are stabilized in a
vacuum with $N=1$, $D=4$ Anti-de~Sitter ($AdS_4$) supersymmetry.
Other superstring theories and more general compactification
schemes will be considered in a longer, companion paper
\cite{dkpzlong}.

This paper is organised as follows. The general method for obtaining
$N=1$ superpotentials from $N=4$ gaugings, already anticipated in
\cite{adk, dkz}, is studied and applied to our specific
compactification scheme in Section~\ref{secgeneral}. The familiar
example of the heterotic theory is then used to define the relation
between fluxes and superpotentials, and the consistency conditions for
a gauging (Section \ref{sechetflux}). We then turn to the general
study of fluxes in type IIA compactifications (Section \ref{secIIA})
and to the study of some selected examples (Section
\ref{secIIAex}). We conclude in Section~\ref{secconc}.

\section{\boldmath{$N=1$} superpotentials from \boldmath{$N=4$} gaugings}
\label{secgeneral}

The Lagrangian density describing the coupling of vector multiplets to
$N=4$, $D=4$ supergravity \cite{N=4} depends on two sets of numbers.
The {\it structure constants} ${f_{ST}}^R$ define the gauge algebra,
and the {\it duality phases} $\delta_R$ specify the duality-covariant
coupling of each gauge field to the supergravity dilaton $S$.  With
$n$ vector multiplets, the gauge group is a $(6+n)$--dimensional
subgroup of the natural $SO(6,n)$ symmetry, inherited from the
superconformal origin of the abelian theory. The structure constants
must leave the $SO(6,n)$ metric $\eta_{RS}$ invariant, a condition
which implies antisymmetry of $f_{STR} \equiv {f_{ST}}^U \eta_{UR}
$. Notice that $\eta_{RU}$ is not in general the Cartan metric of the
gauge group~\footnote{The $SO(6,n)$ metric has six eigenvalues $-1$
and $n$ eigenvalues $+1$.}. With the $SU(1,1) / U(1)$ K\"ahler
potential
\beq
\label{KSis}
K (S,\ov{S}) = - \ln \ (S+\ov{S}) \, ,
\eeq
the $S$--dependent $N=4$ (superconformal) gauge kinetic terms read
\beq
\label{dual2}
{\cal L}_{gauge} =
-{1\over4} \, \sum_{R,S}\,\eta_{RS} \, S_{\delta_R}
F_{\mu\nu}^{R-} F^{\mu\nu\, S- } +  {\rm h.c.} + \ldots \, ,
\eeq
where $F_{\mu\nu}^{R\pm} = F^R_{\mu\nu} \pm i\widetilde F^R_{\mu\nu}$,
and
\beq
\label{dual3}
S_{\delta_R} = \displaystyle{{\rm cos}\,\delta_R
\,S -i\,{\rm sin}\,\delta_R \over
-i\,{\rm sin}\,\delta_R \, S + {\rm cos}\,\delta_R} \, .
\eeq
Further gauge kinetic terms, depending on the scalars in the vector
multiplets, arise from the elimination of superconformal auxiliary
fields  \cite{N=4}. They also depend on the duality phases through the same
$S_{\delta_R}$.  The duality phases $\delta_R$ must respect the
structure of the gauge algebra and discretization of the $SU(1,1)$
$S$--duality group implies that only two choices of phases are
allowed,
\be
\label{delch}
\delta_R=0 \; \leftrightarrow \; S_{\delta_R} = S \, ,
\qquad {\rm and} \qquad
\delta_R=\pi/2 \; \leftrightarrow \; S_{\delta_R} = 1/ S \, ,
\ee
commonly associated with perturbative and non-perturbative sectors,
respectively.

The $n$ vector multiplets contain scalars in the representation {\bf6}
of the $R$--symmetry group $SU(4)$. They live \cite{Chams, DF} on the
coset $SO(6,n) /[ SO(6) \times SO(n)]$:
\beq
\label{scal1}
\phi_{ij}^R = -\phi_{ji}^R = {1\over2} \epsilon_{ijkl}
\phi^{kl\,R} \, , \quad
\phi^{ij\,R} = (\phi_{ij}^R)^* \, ,
\quad
(i,j,\ldots=1,\ldots,4 ,
\quad
R=1,\ldots,6+n) \, .
\eeq
The structure of the sigma-model is dictated by the field equation of
an auxiliary scalar, which leads to the constraint
\beq
\label{scal2}
\eta_{RS}\,\phi_{ij}^R\phi^{kl\,S} = {1\over12}\left(
\delta_i^k \delta_j^l -   \delta_i^l \delta_j^k  \right)
\eta_{RS} \, \phi_{mn}^R\phi^{mn\,S} \, ,
\eeq
and by the Poincar\'e gauge-fixing condition
\beq
\label{scal3}
\eta_{RS}\,\phi_{ij}^R\phi^{ij\,S}
\equiv \phi_{ij}^R\phi^{ij}_R = - 6 \, .
\eeq
These two conditions eliminate twenty-one scalar fields, and the local
$SU(4)$ symmetry can be used to eliminate another fifteen. The
remaining $6n$ physical scalars live on the announced coset.

As usual, gauging supergravity also generates a scalar potential, and
gravitino mass terms $- (1/2) \, {{\cal M}_{3/2}}^{ij} \, \ov\psi_{\mu
i} \sigma^{\mu\nu} \psi_{\nu j} + {\rm h.c.}$, with~\footnote{For
$N=4$, $D=4$ supergravity, we mostly follow the conventions of
\cite{WPhD}, unless otherwise stated, and set the $D=4$ Planck mass
equal to one.}
\beq
\label{3/2mass}
{{\cal M}_{3/2}}^{ij} = - \, {4\over3} \,
\varphi_{(R)}^* \, f_{RST} \, \phi^{ikR}\phi_{kl}^S\phi^{ljT} \, ,
\eeq
and
\be
\varphi_{(R)}^* =
\sqrt{2\over S+\ov S} \, \left( {\rm cos}\, \delta_R
-  iS\,{\rm sin}\, \delta_R  \right) \, .
\ee
To reduce supersymmetry to $N=1$, we use a $Z_2\times Z_2$ truncation,
as in string orbifolds with the same discrete point
group. This truncation leads to a moduli sector with seven chiral
multiplets $S, T_A, U_A$, $(A=1,2,3)$, for all string compactifications
and compatible orientifolds and $D$--brane systems. We can also include
an arbitrary number of matter multiplets, generically denoted by
$Z_A^I$, $(I=1,\ldots,n_A)$. The $N=4$ sigma-model reduces to the
K\"ahler manifold
\beq
\label{manif1}
M_{Z_2 \times  Z_2}={SU(1,1)\over U(1)}
\times  \prod_{A=1}^3 {SO(2,2+n_A)\over SO(2)\times SO(2+n_A)} \, .
\eeq
Since
\be
{SO(2,2)\over SO(2)\times SO(2)} = {SU(1,1)
\over U(1)} \times {SU(1,1)\over U(1)} \, ,
\ee
in the absence of further $Z_A^I$ fields each complex modulus is
associated to an $SU(1,1)/U(1)$ structure. In the Lagrangian, the
truncation is performed by first rewriting the fields in an $SU(3)$
basis,
\beq
\label{phiN=1}
\phi^{R\,A} \equiv \phi^{R\,A4} \, ,  \qquad \qquad
\phi^R_A = (\phi^{R\,A})^* = {1\over2}\epsilon_{ABC}\phi^{R\,BC}\, .
\eeq
The three $SU(3)$ non-singlet gravitino multiplets are then truncated,
and the remaining $N=1$ gravitino mass term reads:
\beq
\label{m32first}
m_{3/2} = -{4\over3} \, \varphi_{(R)}^* \, f_{RST}
\, \epsilon_{ABC} \,
\phi^{R\,A} \phi^{S\,B} \phi^{T\,C} \, .
\eeq
This simple formula still depends on the constrained $N=4$ scalar fields
$\phi^{R\,A}$. However, once written in terms of the unconstrained fields,
the expression of the gravitino mass term will considerably change (see
below). These constrained states are truncated to $N=1$ multiplets
according to the $Z_2\times Z_2$ action on the $SU(3)$ and $SO(6,n)$
indices $A$ and $R$, as in the sigma model truncation (\ref{manif1}).
Since our goal is to work with a fixed set of well-defined moduli and
matter fields $(T_A, U_A, Z^I_A)$, and to study various classes of
gaugings of these multiplets, the next step is to solve the truncated
constraints (\ref{scal2}) and (\ref{scal3}). We then introduce three
sets of $4+n_A$ complex scalars that we denote by
\be
\sigma_A^1, \,\, \sigma_A^2, \,\,
\rho_A^1, \,\, \rho_A^2, \,\, \chi^I_A, \qquad\qquad
A=1,2,3, \qquad I=1,.\ldots,n_A \, .
\ee
The truncated, $SO(2,2+n_A)$--invariant constraints, which
for $\eta_{RS} = diag \, ( - 1_6 , \, 1_n )$ read
\beq
\label{Poin1}
\begin{array}{rcl}
|\sigma_A^1|^2 + |\sigma_A^2|^2 -  |\rho_A^1|^2
-  |\rho_A^2|^2 - \sum_I |\chi^I_A|^2
&=& 1/2 \, ,
\crbig
(\sigma_A^1)^2 + (\sigma_A^2)^2
- (\rho_A^1)^2 - (\rho_A^2)^2 - \sum_I (\chi^I_A)^2
&=& 0  \, ,
\end{array}
\eeq
are then solved in this basis by:
\beq
\label{solution}
\begin{array}{rclrcl}
\sigma_A^1 &=& \displaystyle{ {1\over2}\,
{1+T_A U_A - (Z_A^I)^2 \over [Y(T_A, U_A, Z_A^I)]^{1/2}} } \, ,
\qquad&\qquad
\sigma_A^2 &=&  \displaystyle{ {i \over2} \,
{T_A  +U_A \over [Y(T_A, U_A, Z_A^I)]^{1/2}} } \, ,
\crbig
\rho_A^1 &=& \displaystyle{{1\over2}\,
{1-T_AU_A + (Z_A^I)^2 \over [Y(T_A, U_A, Z_A^I)]^{1/2}} } \, , &
\rho_A^2 &=&  \displaystyle{ {i\over2} \, {T_A-U_A
\over [Y(T_A, U_A, Z_A^I)]^{1/2}}} \, ,
\crbig
\chi_A^I &=& i \, \displaystyle{Z_A^I
\over [Y(T_A, U_A, Z_A^I)]^{1/2}} \, . &&&
\end{array}
\eeq
These expressions depend on the real quantity
\beq
\label{Yis}
Y(T, U, Z^I) = (T+\ov T)(U+\ov U)
- \sum_I(Z^I+\ov Z^I)^2 \, .
\eeq
As expected, the constraints eliminate six complex scalar fields.

The above equations allow to rewrite the scalar potential and the
gravitino mass term as functions of the $N=1$ complex scalars, the
structure constants and the duality phases. The K\"ahler potential and
the superpotential can then be obtained by separating the holomorphic
part in the $N=1$ gravitino mass term, using the relation $m_{3/2} =
e^{K/2} W$. The resulting K\"ahler potential is
\beq
\label{Kis}
K = - \ln (S+\ov S) -\sum_{A=1}^3 \, \ln Y(T_A, U_A, Z^I_A) \, ,
\eeq
while the superpotential is simply
\beq
\label{Wis1}
W = {4\over3} \sqrt2 \, \left[ {\rm cos}\,\delta_R - i \,{\rm sin}\,\delta_R S
\right] \biggl[\prod_{A=1}^3  Y(T_A,U_A,Z_A^I) \biggr]^{1/2}
f_{RST} \, \epsilon_{ABC}\, \phi^{R\,A}\phi^{S\,B} \phi^{T\,C} \, .
\eeq
It is a holomorphic function of $(S, T_A, U_A, Z^I_A)$, once the $N=4$
scalars from the vector multiplets have been truncated to $N=1$ and
replaced by the solutions (\ref{solution}).

In this paper, we discard all matter fields $Z^I_A$ for
simplicity. However, many of the features encountered in the
restricted cases studied here remain true with all matter fields
included. Removing the $Z^I_A$ fields, the generic superpotential is
then a polynomial in the moduli fields with maximal degree seven. In
particular, each monomial is of order zero or one in each of the seven
moduli $S$, $T_A$, $U_A$. The superpotential can then have up to
$2^7=128$ real parameters, which are structure constants and duality
phases of the underlying $N=4$ algebra~\footnote{The $N=1$ truncation
of the scalar fields $\phi^{R\,A}$ associates to each fixed value of
$A=1,2,3$ only four values of the index $R$, the four directions in
each of the three $SO(2,2)$. Hence $f_{RST}$ includes $4^3=64$ real
numbers.}. These numbers will be identified with various fluxes of
compactified string theories.

The structure constants ${f_{RS}}^T$ gauge a subalgebra of $SO(6,6)$,
with dimension equal or less than twelve and compatible with the
$Z_2\times Z_2$ truncation. They verify Jacobi identities. The gauging
structure constants with lower indices $f_{RST} = {f_{RS}}^Q
\eta_{QT}$ are fully antisymmetric for consistency of the gauging.
The truncation to $N=1$ provides further information. The residual
Poincar\' e gauge fixing conditions solved by $T_A$ and $U_A$ are
invariant under $SO(3)$ rotations of the plane index $A$ and $SO(2,2)$
rotations inside each plane. This means that structure constants can
be classified using the $SO(2,2)\times SO(3)$ subgroup of $SO(6,6)$
with embedding ${\bf12} = ({\bf4},{\bf3})$. One can rewrite the gauge
algebra in this embedding by defining generators $T_{Aa}$, ($A=1,2,3,
\, a=1,\ldots,4$), and commutation relations
\beq
\label{gauging1}
[ T_{Aa} , T_{Bb} ] = {f_{Aa\,Bb}}^{Cc} \, T_{Cc} \, .
\eeq
The antisymmetric gauging structure constants are then
\beq
\label{gauging2}
f_{ Aa \, Bb \, Cc } =   {f_{Aa\,Bb}}^{Cd}  \eta_{cd} \, ,
\eeq
where $ \eta_{cd}$ is the $SO(2,2)$ metric. The $Z_2\times Z_2$
orbifold projection~\footnote{And the $SO(3)$ invariance of the
constraints.} leads naturally to define a `plane-interchange symmetry'
in the moduli sector. Our purpose here is to study a particular class
of gaugings which respect this plane-interchange symmetry. Ref.
\cite{dkpzlong} will analyse more general gauging structures.  The
structure constants for these particular gaugings read
\beq
\label{gauging3}
\begin{array}{l}
f_{Aa_1 \, Bb_2 \, Cc_3} = \Lambda_{a_1b_2c_3} \, \epsilon_{ABC}
\, , \qquad \qquad
(a_1,b_2,c_3 = 1, \ldots, 4) \, ,
\crbig
{f_{Aa_1\,Bb_2}}^{Cc_3} =  {\Lambda_{a_1b_2}}^{c_3} \, \epsilon_{ABC}
\, , \qquad \qquad
{\Lambda_{a_1b_2}}^{c_3} = \eta^{c_3d_3}\Lambda_{a_1b_2d_3} \, .
\end{array}
\eeq
Each index $a_1,b_2,c_3$ is an $SO(2,2)$ index, and there are in
principle $4^3=64$ possible combinations, as for the number of
possible superpotential terms constructed with $U_A$ and $T_A$ and the
rule that each term is either linear or independent of each modulus
($2^6=64$). Each $SO(2,2)$ index refers to a specific complex plane of
the $N=1$ truncation: $a_1$ to the first plane, $b_2$ to the second,
$c_3$ to the third. Antisymmetry of the gauging structure constants
$f_{RST}$ implies full symmetry of $\Lambda_{a_1b_2c_3}$: this
reduces the number of independent structure constants to 20, which is
also the number of combinations of superpotential terms left invariant
by any permutation of the plane index. The Jacobi identities verified
by the structure constants ${f_{Aa \, Bb}}^{Cc}$ translate into a
simple cyclicity property:
\beq
\label{gauging4}
\eta^{df} \Lambda_{abd} \Lambda_{cfe} = \eta^{df}\Lambda_{bcd}
\Lambda_{afe} = \eta^{df}\Lambda_{cad} \Lambda_{bfe} \, , \qquad\qquad
\forall a, b, c, e .
\eeq
Eq. (\ref{gauging4}) and symmetry of $\Lambda_{abc}$ are the
conditions applying to an $N=4$ gauging respecting the
plane-interchange symmetry.

There are two commonly used bases for $SO(6,6)$. Firstly, the natural
basis in which the Cartan metric is diagonal, as in eq.~(\ref{Poin1}).
Secondly, the S/A basis defined by
\be
ds^2 = \sum_{i=1}^6 [dx^{i+}dx^{i+} - dx^{i-}dx^{i-} ] = \sum_{i=1}^6
(dx^{i+} + dx^{i-})(dx^{i+} - dx^{i-}) \equiv
\sum_{i=1}^6 dx^{is} dx^{ia} \, .
\ee
The Cartan metric in the S/A basis is off-diagonal,
\beq
\label{CartanSA}
\eta  = {1\over2} \left(\begin{array}{cc} 0_6 & I_6 \\ I_6 & 0_6
\end{array}\right).
\eeq
The analysis of the consistency conditions (\ref{gauging4}) is much
simpler in the S/A basis, especially in view of the solutions
(\ref{solution}) of the Poincar\'e constraints.  The procedure to
analyse a gauging and obtain the corresponding $N=1$ superpotential is
as follows. To begin with, select a symmetric set of constants
$\Lambda_{a_1 b_2 c_3}$ that solve the cyclicity equations
(\ref{gauging4}) in the S/A basis.  Then, compute the resulting $N=1$
superpotential, in two steps. Firstly, use the following
correspondence between the indices ($a_1, b_2, c_3$) and the twelve
directions in $SO(6,6)$:
\beq
\label{gauging21}
\begin{array}{rcll}
a_1 &=& (1,2,3,4) \qquad&\leftrightarrow\qquad (5_S, 6_S, 5_A, 6_A)
\, ,
\crbig
b_2 &=& (1,2,3,4) \qquad&\leftrightarrow\qquad (7_S, 8_S, 7_A, 8_A)
\, ,
\crbig
c_3 &=& (1,2,3,4) \qquad&\leftrightarrow\qquad (9_S, 10_S, 9_A, 10_A)
\, .
\end{array}
\eeq
Secondly, use the solution of the Poincar\'e constraints, in the form
\beq
\label{gauging22}
\begin{array}{rl}
(5_S, 7_S, 9_S) \qquad&\rightarrow\qquad 1/\sqrt{(T_A+\ov T_A)(U_A+\ov
U_A)}, \qquad A=1,2,3 \, , \crbig
(6_S, 8_S, 10_S) \qquad&\rightarrow\qquad iT_A /\sqrt{(T_A+\ov
T_A)(U_A+\ov U_A)}, \qquad A=1,2,3 \, , \crbig
(5_A, 7_A, 9_A) \qquad&\rightarrow\qquad T_AU_A/\sqrt{(T_A+\ov
T_A)(U_A+\ov U_A)}, \qquad A=1,2,3 \, , \crbig
(6_A, 8_A, 10_A) \qquad&\rightarrow\qquad iU_A/\sqrt{(T_A+\ov
T_A)(U_A+\ov U_A)}, \qquad A=1,2,3 \, .
\end{array}
\eeq
For each compactified string theory, the allowed fluxes will determine
the set of allowed $\Lambda_{a_1b_2c_3}$ and the cyclicity equations
(\ref{gauging4}) will impose the consistency relations between various
fluxes. This method can be used to generate all superpotentials from
fluxes verifying plane-interchange symmetry.  Without invoking this
symmetry, the analysis of a gauging would be similar, but with a set
of nonzero gauging structure constants submitted to more complicated
Jacobi identities, instead of the simple relations (\ref{gauging4}).

With our seven moduli fields, K\"ahler potential (\ref{Kis})
and superpotential (\ref{Wis1}), the $N=1$ supergravity scalar potential
simplifies to
\beq
\label{pot3}
e^{-K} V = \sum_{i=1}^7 |W - W_i(z_i + \ov z_i)|^2 - 3 |W|^2,
\eeq
where $z_i = S, T_A, U_A$ and $W_i=(\partial W)/(\partial z_i)$.
Each quantity $[W - W_i(z_i + \ov z_i)]$ is simply the superpotential
$W$ with the corresponding field $z_i$ replaced by $-\ov z_i$.


\section{Heterotic fluxes}
\label{sechetflux}

Before moving to the discussion of the IIA theory, we recall some
known results for $N=1$ compactifications of the heterotic theory on
the $T^6/ (Z_2 \times Z_2)$ orbifold. This will be useful to establish
some notation and to illustrate our general method in a familiar case.

We begin with the identification of the seven main moduli.
Conventionally, we split the space-time indices as $M = [\mu = 0, 1,
2, 3; \ i = 5, 6, 7, 8, 9, 10]$, and we take one $Z_2$ acting on the
coordinates $x^{5, 6, 7, 8}$, the other $Z_2$ on the coordinates
$x^{7, 8, 9, 10}$. This naturally defines three complex planes
$A=1,2,3$: $i_1=5,6$, $i_2=7,8$, $i_3=9,10$. We follow the conventions
of \cite{polbook} unless otherwise stated.

If we neglect the $E_8 \times E_8$ or $SO(32)$ gauge bosons (which
would generate multiplets of $Z_A^I$ type in the $D=4$ theory, and
would allow for additional fluxes associated with the internal
components of their two-form field strengths), the bosonic fields of
the $D=10$ heterotic theory are just the universal ones of the
NS-NS sector: the string-frame metric $g_{MN}$, the
dilaton $\Phi$ and the two-form potential $B_{MN}$. Their $Z_2 \times
Z_2$ invariant components can be decomposed as:
\be
e^{\dd - 2 \Phi} = s \ (t_1 \, t_2 \, t_3)^{- 1} \, ,
\qquad
g_{\mu \nu} = s^{-1} \ \widetilde{g}_{\mu \nu} \, ,
\label{dilmet}
\ee
\be
\label{tudef}
g_{i_A j_A} =
{t_A \over u_A} \ \left(
\begin{array}{cc}
u_A^2 + \nu_A^2 & \nu_A \\
\nu_A & 1
\end{array} \right) \, ,
\quad
(A=1,2,3) \, ,
\ee
\be
B_{\mu \nu} \leftrightarrow \sigma \, ,
\quad
B_{56} = \tau_1 \, ,
\quad
B_{78} = \tau_2 \, ,
\quad
B_{910} = \tau_3 \, ,
\label{sigmataudef}
\ee
where $\widetilde{g}_{\mu \nu}$ is the metric in the $D=4$ Einstein
frame, and the symbol $\leftrightarrow$ indicates the four-dimensional
duality transformation relating a two-form potential with an axionic
pseudoscalar. Neglecting the dependence of the fields on the internal
coordinates, absorbing an integration constant in the $D=4$ Planck
mass, conventionally set to unity, and making the identifications
\be
S = s + i \, \sigma \, ,
\quad
T_A = t_A + i \, \tau_A \, ,
\quad
U_A = u_A + i \, \nu_A \, ,
\quad
(A=1,2,3) \, ,
\label{studef}
\ee
we obtain $D=4$ kinetic terms described precisely by the K\"ahler
potential of eq.~(\ref{Kis}), for the case $Z_A^I=0$ we have chosen to
study:
\be
\label{kahler}
K = - \ln \, (S + \ov{S})
- \sum_{A=1}^3 \ln \, (T_A + \ov{T}_A)
- \sum_{A=1}^3 \ln \, (U_A + \ov{U}_A) \, .
\ee
In view of what follows, we stress that the kinetic terms of the seven
main moduli are invariant under both $O(7)$ rotations and $SU(1,1)
\times [SO(2,2)]^3$ duality transformations.

We now summarize the different allowed fluxes, and identify the
associated $N=1$ superpotentials with the method illustrated in the
previous section.

\subsection{\boldmath{$\widetilde{H}_3$} heterotic fluxes}

As first recognized in \cite{h3het}, possible fluxes in the heterotic
theory are those of the modified NS-NS three-form $\widetilde{H_3} =
dB_2 + \ldots$, where the dots stand for the gauge and Lorenz
Chern-Simons terms. There are eight independent real fluxes, invariant
under the $Z_2 \times Z_2$ orbifold projection:
\be
\label{h3fluhet}
\widetilde{H}_{579} \, ,
\;
\widetilde{H}_{679} \, ,
\;
\widetilde{H}_{589} \, ,
\;
\widetilde{H}_{689} \, ,
\;
\widetilde{H}_{5710} \, ,
\;
\widetilde{H}_{6710} \, ,
\;
\widetilde{H}_{5810} \, ,
\;
\widetilde{H}_{6810} \, .
\ee
The corresponding potential for the seven main moduli can be
explicitly computed by dimensional reduction. Its generic structure is
$V_{H_3} = e^K \, \prod_{A=1}^3 f_A ( \nu_A, u_A^2 + \nu_A^2 )$, where
each $f_A$ is a polynomial of at most degree one in its
arguments. This is sufficient to deduce, similarly to what happens in
IIB theories \cite{IIBflu}, the corresponding $N=1$ effective
superpotential $W_{H_3} (U)$, which carries no dependence on the $S$
and $T$ moduli. It is immediate to check that $\widetilde{H}_3$ fluxes
correspond to $N=4$ gaugings for any choice of the parameters in
(\ref{h3fluhet}). Leaving aside a systematic discussion, we just
observe that, under the assumption of plane-interchange symmetry,
there are four independent parameters, associated with four different
structures in $W_{H_3}(U)$:
\bea
\widetilde{H}_{579} \equiv \Lambda_{111} & 
\leftrightarrow & 1 \, , \nn \\
\widetilde{H}_{679} = \widetilde{H}_{589} = 
\widetilde{H}_{5710} \equiv \Lambda_{114}
& \leftrightarrow & i \ (U_1 + U_2 + U_3) \, , \nn \\
\widetilde{H}_{689} = \widetilde{H}_{5810} = 
\widetilde{H}_{6710} \equiv \Lambda_{144} & \leftrightarrow
& - (U_1 \, U_2 + U_2 \, U_3 + U_1 \, U_3) \, , \nn \\
\widetilde{H}_{6810} \equiv \Lambda_{444} & \leftrightarrow & - i
\ U_1 \, U_2 \, U_3 \; .
\label{pish3}
\eea

\subsection{Geometrical heterotic fluxes}

The possible fluxes also include some geometrical ones, associated
with the internal components of the spin connection $\omega_3$, and
corresponding to coordinate-dependent compactifications
\cite{ftss}. These fluxes are characterized by real constants with one
upper curved index and two lower antisymmetric curved indices:
\be
f^i_{\;\; j k} = - f^i_{\;\; k j} \, .
\ee
These constants must satisfy the Jacobi identities of a Lie group,
$f^{i}_{\;\; j k} \ f^{k}_{\;\; l m} + f^{i}_{\;\; l k} \ f^{k}_{\;\;
m j} + f^{i}_{\;\; m k} \ f^{k}_{\;\; j l} = 0$, and the additional
consistency condition $f^{i}_{\;\; i k} = 0$. The corresponding $D=4$
potential in the heterotic theory can be easily calculated from the
formulae in \cite{ftss}.

In agreement with the $Z_2 \times Z_2$ orbifold projection, we
must assume here that
\be
f^{i_A}_{\quad i_B i_C} = 0 \qquad {\rm for} \;\;\;\;
A=B \;\; {\rm or} \;\; A=C \; {\rm or} \;\; B=C \, ,
\label{parss1}
\ee
which satisfies automatically the consistency condition $f^{i}_{\;\;
i k}=0$. Geometrical fluxes are then described by 24 real parameters,
\be
C_{i_A i_B i_C} \equiv f^{i_A}_{\quad i_B i_C} \, ,
\qquad
[(ABC)=(123),(231),(312)] \, ,
\label{parss2}
\ee
subject only to the Jacobi identities. Leaving aside a general
discussion, we assume here plane-interchange symmetry, to reduce the
number of independent parameters. Inspection of the resulting scalar
potential singles out six different possible structures in the
effective superpotential $W_{\omega} (T,U)$, always linear in the $T$
moduli and independent of $S$:
\bea
C_{679} = C_{895} = C_{1057} \equiv \Lambda_{112}
& \leftrightarrow & i \ (T_1 + T_2 + T_3) \, , \nn \\
C_{579} = C_{957} = C_{795} \equiv \Lambda_{113}
& \leftrightarrow & (T_1 \, U_1 + T_2 \, U_2 +
T_3 \, U_3) \, , \nn \\
C_{6810} = C_{8106} = C_{1068} \equiv \Lambda_{244}
& \leftrightarrow & - i \ (T_1 \, U_2 \, U_3 + T_2 \, U_1
\, U_3 +  T_3 \, U_1 \, U_2) \, , \nn \\
C_{5810} = C_{7106} = C_{968} \equiv \Lambda_{344}
& \leftrightarrow &  - (T_1 + T_2 +  T_3) \,
U_1 \, U_2 \, U_3 \, , \nn \\
\begin{array}{l} C_{896} = C_{1067} = C_{689} = \\
C_{1058} = C_{6710} = C_{8105} \equiv \Lambda_{124}
\end{array}
& \leftrightarrow & \begin{array}{c}
- (T_1 U_2 + T_1 U_3 + T_2 U_1 \\ +
T_2 U_3 +  T_3 U_1 + T_3 U_2) \end{array}
\, , \nn \\
\begin{array}{l} C_{589} = C_{796} = C_{7105} = \\
C_{958} = C_{5710} = C_{967} \equiv \Lambda_{134}
\end{array}
& \leftrightarrow & \ \begin{array}{c}
i \ (T_1 U_1 U_2 + T_2 U_2 U_3 + T_3 U_3 U_1 \\
+T_1 U_1 U_3 + T_2 U_2 U_1 + T_3 U_3 U_2)
\end{array} \, .
\label{pisss}
\eea
In this case the Jacobi identities (\ref{gauging4}) impose some
non-trivial constraints:
\bea
\Lambda_{112} \, \Lambda_{344} & = &
\Lambda_{124} \, \Lambda_{134} \, ,
\nn \\
\Lambda_{113} \, \Lambda_{344}
+
\Lambda_{244} \, \Lambda_{134}
& = &
\Lambda_{344} \, \Lambda_{124}
+
\Lambda_{134}^2 \, ,
\nn \\
\Lambda_{112} \, \Lambda_{244}
+
\Lambda_{113} \, \Lambda_{124}
& = &
\Lambda_{112} \, \Lambda_{134}
+
\Lambda_{124}^2 \, .
\label{geohet}
\eea
\subsection{Combined heterotic fluxes}
The combination of $\widetilde{H}_3$ and $\omega_3$ fluxes in the
$T^6/(Z_2 \times Z_2)$ orbifold of the heterotic string is then
described, under the assumption of plane-interchange symmetry, by the
ten real parameters of eqs.~(\ref{pish3}) and (\ref{pisss}). According
to eq.~(\ref{gauging4}), consistency with an underlying $N=4$ gauging
amounts to requiring the Jacobi identities (cyclicity conditions) of
eq.~(\ref{geohet}) and the additional condition
\be
\Lambda_{111} \, \Lambda_{344}
+
\Lambda_{112} \, \Lambda_{444}
+
\Lambda_{113} \, \Lambda_{144}
+
\Lambda_{114} \, \Lambda_{244}
=
2 \, \Lambda_{144} \, \Lambda_{124}
+
2 \, \Lambda_{114} \, \Lambda_{134} \, .
\label{addhet}
\ee
The corresponding effective $N=1$ superpotential would read
\bea
W & = & \Lambda_{111}
+ i \, \Lambda_{114} \ (U_1 + U_2 + U_3)
- \Lambda_{144} \ (U_1 U_2 + U_2 U_3 + U_1 U_3)
- i \  \Lambda_{444} \ U_1 U_2 U_3
\nn \\ & &
+ i \, \Lambda_{112} \ (T_1 + T_2 + T_3)
+ \Lambda_{113} \ (T_1 U_1 + T_2 U_2 + T_3  U_3)
\nn \\ & &
- i \, \Lambda_{244} \ (T_1 U_2 U_3 + T_2 U_1 U_3 + T_3 U_1 U_2)
- \Lambda_{344} \ (T_1 + T_2 + T_3) \ U_1 U_2 U_3
\nn \\ & &
- \Lambda_{124} \ (T_1 U_2 + T_1 U_3 + T_2 U_1
+ T_2 U_3 +  T_3 U_1 + T_3 U_2)
\nn \\ & &
+ i \ \Lambda_{134} \ (T_1 U_1 U_2 + T_1 U_1 U_3 + T_2 U_2 U_1
+ T_2 U_2 U_3 +  T_3 U_3 U_1 + T_3 U_3 U_2) \, .
\label{hetsup}
\eea
Similar superpotentials were considered, motivated by $N>1$ gaugings
but without assuming plane interchange symmetry and without
establishing the precise connections with fluxes, in \cite{fkz}.  The
connection between Scherk-Schwarz compactifications, geometrical
fluxes, $N=4$ gaugings and $N=1$ superpotentials was also discussed in
\cite{pz}, without assuming plane-interchange symmetry and in a
different field basis. More results at the $N=4$ level were obtained
in \cite{ssvsn4}. A general analysis of combined fluxes in toroidal
compactifications of the heterotic string was given in \cite{kalmy}:
consistent $Z_2 \times Z_2$ truncations of their results are in
complete agreement with our results.

It is important to recall that, in the heterotic theory,
$\widetilde{H}_3$ and $\omega_3$ fluxes, corresponding to perturbative
$N=4$ gaugings with trivial duality phases, can never generate $N=1$
superpotentials with both constant and linear terms in $S$. We can
then obtain, for example, no-scale models as in \cite{strss, pz}, but
never reach the full stabilization of all seven main moduli, including
$S$. From the point of view of $N=4$ supergravity, of course, we could
also consider non-perturbative gaugings with non-trivial duality
phases, which would give rise to both kinds of allowed $S$-dependences
in the effective $N=1$ superpotential. We may think of these gaugings
as associated to possible non-perturbative effects such as gaugino
condensation.


\section{Fluxes in IIA superstrings}
\label{secIIA}
In type IIA and IIB superstring theories compactified on $T^6/(Z_2
\times Z_2)$, to produce $N=1$, $D=4$ supersymmetry we must introduce
consistently an additional $Z_2$ orientifold projection. We discuss
here only the case of the IIA theory, with a specific orientifold
projection compatible with D6 branes.

The bosonic fields of the IIA theory are the universal ones of the
NS-NS sector, plus those of the R-R sector: a one-form $A_{M}$ and a
three-form $A_{MNR}$. Five-forms and seven-forms are related to the
previous ones by ten-dimensional duality, do not carry independent
degrees of freedom and do not need to be included at this
stage. Nine-form potentials do not carry any propagating degree of
freedom in $D=10$, even if they can play a role, as we shall see, in
the classification of allowed fluxes. We consider here a specific
$Z_2$ orientifold projection, involving the inversion of three out of
the six internal coordinates and associated with D6 branes: it is not
restrictive to take the odd coordinates to be $x^{5,7,9}$. The
independent invariant spin-0 fields from the NS-NS sector are then:
\be
g_{i i} \, ,
\quad
\Phi \, ,
\quad
B_{5 6} \, ,
\quad
B_{7 8} \, ,
\quad
B_{9 10} \, ,
\ee
for which we can temporarily make the heterotic decomposition of
eqs.~(\ref{dilmet})--(\ref{sigmataudef}), setting $\nu_A \equiv 0$ and
disregarding the off-diagonal components of the internal metric. The
independent invariant spin-0 fields from the R-R sector are:
\be
A_{6 8 10} = \sigma^{\, \prime} \, ,
\quad
A_{6 7 9} = - \nu_1^{\, \prime} \, ,
\quad
A_{5 8 9} = - \nu_2^{\, \prime} \, ,
\quad
A_{5 7 10} = - \nu_3^{\, \prime} \, .
\label{ramIIA}
\ee
Looking at the $D=4$ kinetic terms of the fields in (\ref{ramIIA}),
we find
\be
{\cal L}_{R} \rightarrow
- {1 \over 4}  \ \widetilde{e}_{4} \
\widetilde{g}^{\mu \nu} \ \left[
{\cal O}_0 \ (\partial_\mu \sigma ^{\, \prime})
\ (\partial_\nu \sigma ^{\, \prime})
+ \sum_{A=1}^3 \ {\cal O}_A \ (\partial_\mu \nu_A^{\, \prime}) \
(\partial_\nu \nu_A^{\, \prime}) \right] \, ,
\label{kinRIIA}
\ee
where
\be
{\cal O}_0 = { u_1 \ u_2 \ u_3 \over s }
\, , \quad
{\cal O}_1 = { u_1 \over s \ u_2 \ u_3 }
\, , \quad
{\cal O}_2 = { u_2 \over s \ u_1 \ u_3 }
\, , \quad
{\cal O}_3 = { u_3 \over s \ u_1 \ u_3 } \, .
\ee
This immediately suggests \cite{aft} the identification of the real
parts $(s^{\, \prime}, u_1^{\, \prime}, u_2^{\, \prime}, u_3^{\,
\prime})$, associated by $N=1$ supersymmetry to the imaginary parts
$(\sigma^{\, \prime}, \nu_1^{\, \prime}, \nu_2^{\, \prime}, \nu_3^{\,
\prime})$:
\be
s^{\, \prime} = \sqrt{s \over u_1 \ u_2 \ u_3} \, ,
\quad
u_1^{\, \prime} = \sqrt{s \ u_2 \ u_3 \over u_1} \, ,
\quad
u_2^{\, \prime} = \sqrt{s \ u_1 \ u_3 \over u_2} \, ,
\quad
u_3^{\, \prime} = \sqrt{s \ u_1 \ u_2 \over u_3} \, .
\label{IIA3red}
\ee
These identifications can be cross-checked by looking at the $F_{\mu
\nu} \, F^{\mu \nu}$ and $F_{\mu \nu} \, \widetilde{F}^{\mu \nu}$
terms in the effective four-dimensional action for the Yang-Mills
vectors, generated by the Dirac-Born-Infeld and Wess-Zumino actions
for the D6 branes, aligned along the $(6810)$, $(679)$, $(589)$,
$(5710)$ O6-planes.

The theory under consideration exhibits a rich structure of possible
invariant fluxes. As in the heterotic case, we first discuss each of
them separately, then we look at a generic combination. The Jacobi
identities of $N=4$ gaugings, eq.~(\ref{gauging4}), will be
automatically satisfied if there are only NS-NS three-form fluxes, or
only fluxes of the R-R forms: as we shall see, non-trivial constraints
will arise only in the presence of geometrical fluxes or of combined
fluxes.

\subsection{\boldmath{$H_3$} fluxes in IIA}

Only four out of the eight independent fluxes allowed for the NS-NS
3-form in the heterotic case, eq.~(\ref{h3fluhet}), are also invariant
with respect to the orientifold projection. In terms of $H_3 = d B_2$,
they are:
\be
\label{3fluxesIIA}
H_{579} \, ,
\quad
H_{689} \, ,
\quad
H_{6710} \, ,
\quad
H_{5810} \, .
\ee
Inspection of the four-dimensional potential obtained from dimensional
reduction shows that, after moving to the IIA field basis of
eq.~(\ref{IIA3red}), and assuming plane-interchange symmetry, there
are two independent superpotential structures:
\bea
H_{579} = \Lambda^{\, \prime}_{111} & \leftrightarrow &
i \ S \, , \nn \\
-H_{689}=-H_{6710}=-H_{5810} = \Lambda_{114} & \leftrightarrow &
i \ (U_1 + U_2 + U_3) \, .
\eea
Notice that, in contrast with the heterotic case, two independent
$SU(1,1)$ phases are involved.

\subsection{Geometrical IIA fluxes}

Again, the IIA orientifold projection leaves invariant only half of
the geometrical fluxes (\ref{parss2}) that were allowed, modulo Jacobi
identities, in the heterotic case:
\be
C_{5710} \, , C_{7105} \, , C_{1057} \, ;
\quad
C_{679} \, , C_{796} \, , C_{967} \, ;
\quad
C_{589} \, , C_{895} \, , C_{958} \, ;
\quad
C_{6810} \, , C_{8106} \, , C_{1068} \, .
\label{parssIIA}
\ee
Inspection of the four-dimensional potential obtained from dimensional
reduction shows that in this case, after moving to the IIA field basis
of eq.~(\ref{IIA3red}), and assuming plane-interchange symmetry, there
are three independent superpotential structures:
\bea
C_{679} = C_{895} = C_{1057} \equiv \Lambda^{\, \prime}_{112}
& \leftrightarrow & - S \ (T_1 + T_2 + T_3) \, , \nn \\
C_{6810} = C_{8106} = C_{1068} \equiv \Lambda_{113}
& \leftrightarrow &  (T_1 \, U_1 + T_2 \, U_2
+  T_3 \, U_3) \, , \nn \\
\begin{array}{l}  C_{589} = C_{796} = C_{7105} = \\
C_{958} = C_{5710} = C_{967} \equiv \Lambda_{124}
\end{array}
& \leftrightarrow & \begin{array}{c}
- (T_1 U_2 + T_1 U_3 + T_2 U_1 \\ +
T_2 U_3 +  T_3 U_1 + T_3 U_2) \end{array}
\, .
\label{pisss2a}
\eea
Notice that also for geometrical fluxes two different $SU(1,1)$ phases
appear, in contrast with the heterotic case. From eq.~(\ref{gauging4})
we can easily derive the Jacobi identities required for a consistent
$N=4$ gauging with geometrical fluxes only:
\be
\Lambda_{124} \ (\Lambda_{124}-\Lambda_{113}) = 0 \, .
\ee

\subsection{\boldmath{$F_0$} flux}
The mass parameter of massive IIA supergravity \cite{romans} can be
regarded as a ten-dimensional zero-form flux $F_0$, dual to the
ten-form field strength associated with a nine-form potential, which
does not carry any propagating degree of freedom. Inspecting the $F_0$
contribution to the potential via dimensional reduction, and moving to
IIA variables, we can identify the associated structure in the
effective superpotential $W$:
\bea
F_0 = \Lambda_{222} & \leftrightarrow & - i \ (T_1 \, T_2 \, T_3) \, .
\eea

\subsection{\boldmath{$F_2$} fluxes}

The independent $F_2$ fluxes invariant under the orbifold and
orientifold projections are:
\be
F_{56} \, ,
\qquad
F_{78} \, ,
\qquad
F_{910} \, .
\label{2fluxesIIA}
\ee
Looking at their contributions to the potential via dimensional reduction,
moving to the IIA field basis of eq.~(\ref{IIA3red}), and assuming
plane-interchange symmetry, we find that the corresponding structure
in the effective superpotential $W$ is
\bea
F_{56}=F_{78}=F_{910} = \Lambda_{122} & \leftrightarrow &
- (T_1 \, T_2 + T_1 \, T_3 + T_2 \, T_3 ) \, .
\eea

\subsection{\boldmath{$F_4$} fluxes}

The four-form fluxes with internal indices, invariant under the
orbifold and orientifold projections, are:
\be
\label{4fluxesIIA}
F_{5678} \, ,
\qquad
F_{78910} \, ,
\qquad
F_{91056} \, .
\ee
Looking at their contribution to the potential via dimensional
reduction, moving to the IIA field basis of eq.~(\ref{IIA3red}),
and assuming plane-interchange symmetry, we can identify the
corresponding structure in the effective superpotential:
\bea
F_{5678}=F_{78910}=F_{56910} = \Lambda_{112} & \leftrightarrow &
i \ (T_1 + T_2 + T_3) \, .
\eea

\subsection{\boldmath{$F_6$} flux}

Among the components of the R-R four-form field strength, invariant
under both the orbifold and the orientifold projections, there is also
$F_{\mu \nu \rho \sigma}$, which is not associated with any $D=4$
propagating degree of freedom, and can be related by ten-dimensional
duality to a ten-dimensional six-form flux $F_6$.  A similar flux was
considered in \cite{boupol} to address the cosmological constant
problem. Looking at the corresponding potential terms generated by
dimensional reduction, and moving to the IIA variables of
eq.~(\ref{IIA3red}), we can identify the corresponding structure in
the effective superpotential:
\bea
F_6 = \Lambda_{111} & \leftrightarrow &  1 \, .
\eea
Notice that the above flux generates a constant superpotential, not a
constant potential, in the four-dimensional effective theory.

\subsection{Combined IIA fluxes}

Switching on simultaneously all the independent fluxes identified so
far corresponds to having, as non-vanishing coefficients:
\begin{itemize}
\item
($ \Lambda_{111}^{\, \prime}, \Lambda_{112}^{\,
\prime}$) with $SU(1,1)$ phase factor $iS$;
\item
($\Lambda_{111}, \Lambda_{112},
\Lambda_{122}, \Lambda_{222}, \Lambda_{113}, \Lambda_{114},
\Lambda_{124}$) with $SU(1,1)$ phase $1$.
\end{itemize}
Under our simplifying assumption of plane-interchange symmetry, the
Jacobi identities constraining such combined fluxes read
\be
\Lambda_{222} \ \Lambda_{114} +
\Lambda_{113} \ \Lambda_{122} =
2 \ \Lambda_{122} \ \Lambda_{124}
\, , \qquad
\Lambda_{113} \Lambda_{124}
= \Lambda_{124}^2 \, .
\label{2ajac}
\ee
Any combination of fluxes satisfying the above Jacobi identities
corresponds to a $N=4$ gauging, and can be easily translated into
an effective $N=1$ superpotential:
\bea
W & = & \Lambda_{111} + i \, \Lambda^{\, \prime}_{111} \ S
+ i \, \Lambda_{112} \ (T_1 + T_2 + T_3)
- \Lambda^{\, \prime}_{112} \ S \ (T_1 + T_2 + T_3)
\nn \\ & &
+ i \, \Lambda_{114} \ (U_1 + U_2 + U_3)
+ \Lambda_{113} \ (T_1 U_1 + T_2 U_2 + T_3  U_3)
- \Lambda_{122} \ (T_1 T_2 + T_1 T_3 + T_2 T_3)
\nn \\ & &
- \Lambda_{124} \ (T_1 U_2 + T_1 U_3 + T_2 U_1
+ T_2 U_3 +  T_3 U_1 + T_3 U_2)
- i \  \Lambda_{222} \ T_1 T_2 T_3 \, .
\label{2asup}
\eea
The results of eqs.~(\ref{2ajac}) and (\ref{2asup}) provide a powerful
and practical tool for analyzing, directly in the $N=1$, $D=4$
effective theory, the different vacuum structures associated with the
different allowed combinations of fluxes, in the chosen orbifold and
orientifold of the IIA theory. The study of a large number of examples
of flux configurations \cite{dkpzlong}, in heterotic and type II
strings, actually shows that the effective supergravity approach based
upon $N=4$ gaugings can accurately reproduce the conditions imposed by
the full field equations of the ten-dimensional theories.  This of
course requires to include all necessary brane and orientifold plane
contributions to these equations.  For the combinations of fluxes
leading to stable vacua of our $N=1$, $D=4$ effective theory, it would
be interesting to explicitly examine the corresponding combinations of
D6-branes and O6-planes required to satisfy the $D=10$ equations and
Bianchi identities, and the associated tadpole cancellation
conditions. This analysis goes beyond the scope of the present paper.


\section{Some selected IIA examples}
\label{secIIAex}
We present now some selected examples of admissible IIA fluxes that
correspond to $N=4$ gaugings with non-trivial $SU(1,1)$ phases and
give rise to physically different situations.

\subsection{Flat gaugings, no-scale models: stabilization of four moduli}

Switching on a system of $(\omega_3,H_3,F_0,F_2)$ fluxes, with nonzero
parameters ($B, \, D>0$)
\be
\begin{array}{ll}
\Lambda_{112}^{\, \prime} = - A \, , 
\qquad
& 
\Lambda_{122} = - A B \, ,
\crbig
\Lambda_{111}^{\, \prime} = C \, , 
\qquad 
& 
\Lambda_{222} = - CD \, ,
\end{array}
\ee
the Jacobi identities (\ref{2ajac}) are automatically satisfied,
and the following effective $N=1$ superpotential is generated:
\be
\label{Wex1}
W = A \Bigl[ S  (T_1 + T_2 + T_3) + B (T_1 T_2 + T_2 T_3 + T_1 T_3) \Bigr]
+ i C \Bigl[ S  + D \, T_1T_2T_3 \Bigr]
\, .
\ee
It is immediate to see that this corresponds to a no-scale model.
Since $W$ does not depend on $(U_1, U_2, U_3)$, the scalar potential
is a sum of positive semi-definite terms,
\beq 
e^{-K} V = |W - (S + \ov S) W_S |^2 +
\sum_{A=1}^3 |W - (T_A+ \ov T_A) W_{T_A} |^2 \, ,
\eeq
and stabilization of the $S$ and $T_A$ moduli occurs at
\beq 
\langle S \rangle = B \, T \, ,
\qquad
\langle T_1 \rangle = \langle T_2 \rangle = 
\langle T_3 \rangle = T \, ,
\qquad 
T = \sqrt{B \over D} \, ,
\eeq
with $\langle V \rangle = 0$. The $U_A$ moduli remain as complex flat
directions and supersymmetry is broken in the $U_A$ sector, since the
stabilization conditions lead to $\langle W \rangle \ne 0$, with a
gravitino mass
\begin{equation}
\langle m_{3/2}^2 \rangle \propto {|9 \, A^2 \, B + C^2 \, D | 
\over u_1 \, u_2 \, u_3} \, .
\end{equation}

To identify the gauging associated with the above fluxes and
superpotential, it is convenient to rescale the fields according to
\begin{equation}
S \,\,\longrightarrow\,\, B^{3/2}D^{-1/2} \, S \, , 
\qquad 
T_A \,\,\longrightarrow\,\, B^{1/2}D^{-1/2} \, T_A \, .
\end{equation}
The stabilization of the rescaled fields occurs at $\langle S \rangle
= \langle T_A \rangle = 1$, with a superpotential as in (\ref{Wex1})
with $B=D=1$. The resulting group is $E_3 \times E_3$ \cite{dkpzlong},
where $E_3$ is the three-dimensional Euclidean group, i.e. the
$SO(3)$--invariant contraction of $SO(4)$ or $SO(3,1)$.

This kind of rescalings can be applied in general, to shift the values
at which the fields are stabilized. For simplicity, and without losing
the full generality of the combination of fluxes, we choose that
moduli are stabilized at value one in most of the following examples.


As a side remark, we notice here that the same phenomenology of the
above example can be obtained without respecting the plane-interchange
symmetry. As an example, we can consider as before a system of fluxes
$(\omega_3, H_3, F_0, F_2)$, but this time corresponding to a 
superpotential:
\begin{equation}
W = A \left(S T_1 + T_2 T_3 \right) + i \, B 
\left(S+T_1T_2T_3\right) \, ,
\quad
{\rm with} 
\quad
\langle m_{3/2}^2 \rangle \propto {A^2 + B^2 \over u_1 \, u_2 \, 
u_3} \, .
\end{equation}
The complex flat directions are $(U_1, U_2, U_3)$, as in the first
example.

We finally notice that, in the type--IIA theory, purely geometrical
fluxes $\omega_3$ are not sufficient to stabilize all moduli
explicitly appearing in the corresponding superpotential $W$, because
the latter is always quadratic in the fields. As an example to
illustrate this point, based on the two-dimensional Euclidean group
$E_2$ [the $SO(2)$--invariant contraction of $SO(3)$ or $SO(2,1)$] and
breaking the plane-interchange symmetry, we consider the
superpotential
\begin{equation}
W = A \, \left( T_1 U_2 + T_2 U_1 \right) \, ,
\quad
{\rm with}
\quad
\langle m_{3/2}^2 \rangle \propto {A^2 \over s \, t_3 \, u_3} \, .
\end{equation}
This corresponds to a $Z_2$ freely acting orbifold (generalized
Scherk-Schwarz mechanism in string theory), with complex flat
directions $(S, T_3, U_3)$. However, there are additional flat
directions, because the auxiliary fields associated with $(T_1, U_1,
T_2, U_2)$ are all set to zero by requiring $\tau_1 = \tau_2 = \nu_1 =
\nu_2 = 0$ and $t_1 \, u_2 = t_2 \, u_1$. In this case the spectrum 
has 4 massive and 3 massless ``axions'', 1 massive and 6 massless
``dilatons''.


\subsection{Gaugings with \boldmath{$V > 0$}, cosmological models}

Examples can be easily found, in which less than four moduli are
stabilized and the potential is always strictly positive-definite,
leading to runaway solutions (in time).

Superpotentials with a single monomial are of course examples where no
modulus gets stabilized. For instance, we can choose the fluxes
$\Lambda_{111}=F_6$, $\Lambda_{222}=F_0$ or $\Lambda^{\, \prime}_{111}
= H_3$, corresponding to
\begin{equation}
W=F_6 \, , 
\qquad  
W= -iF_0 \, T_1 T_2 T_3  
\qquad {\rm or} \qquad
W=i H_3 S \, .
\end{equation}
This leads to $V=4 \, e^K \, |W|^2$, with $|W|>0$ and a gravitino mass
term of the form
\begin{equation}
m_{3/2}^2  = {1 \over 2^7 \, s \,
t_1 \, t_2 \, t_3 \, u_1 \, u_2 \, u_3} \, \times \Bigl\{
|F_6|^2 \,\, , \,\, \,
|F_0 T_1 T_2 T_3|^2 \,\,
{\rm or}
\,\, |H_3S|^2
\Bigr\} \, , 
\end{equation} 
respectively.

An example where three moduli are stabilized is obtained by switching
on a system of R-R fluxes $(F_0,F_2,F_4,F_6)$, with parameters
\be 
\Lambda_{111} = - \Lambda_{122} = A \, , 
\qquad 
\Lambda_{112} = - \Lambda_{222} = B \, .
\ee
The Jacobi identities (\ref{2ajac}) are automatically satisfied,
and the following effective $N=1$ superpotential is generated:
\begin{equation}
W = A \left(1 + T_1 T_2 +T_2 T_3 + T_3 T_1\right) + i
B\left(T_1+T_2+T_3+T_1T_2T_3\right) \, .
\end{equation}
This choice of fluxes and superpotential is actually a gauging of
$SO(1,3)$. It is immediate to see that, since the superpotential does
not depend on four of the seven main moduli (the $T$-moduli are
stabilized at one), supersymmetry is broken and a positive-definite
runaway $D=4$ scalar potential is generated,
\begin{equation}
\langle V \rangle = 
\langle m^2_{3/2} \rangle \, ,
\qquad 
{\rm with} 
\qquad 
\langle m^2_{3/2} \rangle 
= {A^2+B^2 \over 8 \, s \, u_1 \, u_2 \, u_3} \, ,
\end{equation}
possibly leading to time-dependent vacua of cosmological interest.

\subsection{Gaugings with \boldmath{$V < 0$}, stabilization of all moduli}

We now look at situations where more than four moduli are stabilized,
leading to negative-definite potentials once the stabilized moduli are
set to their appropriate values.

We begin with a gauging of $E_3$ with fluxes $\Lambda_{113}= -
\omega_3$ (geometric) and $\Lambda_{111} = F_6$ (R-R six-form), with
$\omega_3 \, , \, F_6 > 0$. The R-R six-form corresponds to the
$SO(3)$ directions in $E_3$ while $\omega_3$ corresponds to the
translations. The superpotential reads
\begin{equation}
W = - \omega_3 \left(T_1 U_1 + T_2 U_2 + T_3 U_3\right) + F_6 \, .
\end{equation}
The six equations for the non-trivial supergravity auxiliary fields
are solved at $\langle \tau_A \rangle = \langle \nu_A \rangle =0$ and
$\langle t_1 \, u_1 \rangle = \langle t_2 \, u_2 \rangle = \langle t_3
\, u_3 \rangle = F_6 / \omega_3$. At these values, $W = - 2 \, F_6$,
and the $s$--dependent scalar potential and gravitino mass term read
\begin{equation}
V   = - 2 \, e^K  \, |W|^2 = - {\omega_3^3 \over 16\, F_6 \, s} 
\,, \qquad\qquad
m_{3/2} ^2 =  - {1\over2} \, V \, .
\end{equation}
At the string level, this is the well-known NS five-brane solution
plus linear dilaton, in the near-horizon limit. The original gauging
is $SU(2)$, combined with translations, which emerge as free actions
at the level of the world-sheet conformal field theory. It is
remarkable that this $E_3$ algebra remains visible at the supergravity
level. It is also interesting that, if we allow extra fluxes, induced
by the presence of fundamental-string sources, we can reach $AdS_3$
background solutions with stabilization of the dilaton. All moduli are
therefore stabilized. This has been studied recently at the string
level \cite{IKP}.

Using all fluxes admissible in IIA, $Z_2\times Z_2$ strings, we can
obtain the stabilization of all moduli in $AdS_4$ space--time
geometry. Switching on all fluxes ($\omega_3, H_3, F_0, F_2,
F_4, F_6$), with parameters
\be
- {1 \over 9} \Lambda_{111} =
- {1 \over 2} \Lambda_{112}^{\, \prime} =
{1 \over 6} \Lambda_{113} =
\Lambda_{122} = A \, ,
\ee
\be
{1 \over 2} \Lambda_{111}^{\, \prime} =
- {1 \over 3} \Lambda_{112} =
{1 \over 2} \Lambda_{114} =
-  {1 \over 5} \Lambda_{222} = B \, ,
\ee
the Jacobi identities (\ref{2ajac}) are satisfied for
\be
6 \, A^2 = 10 \, B^2 \, ,
\label{cond}
\ee
and the following effective $N=1$ superpotential is generated:
\bea
W & = &
A \left[ 2 \ S \ (T_1+T_2+T_3) - (T_1 T_2 + T_2 T_3 + T_3T _1 )
+ 6 \ (T_1 U_1 +T_2 U_2 + T_3 U_3 ) - 9 \right] \nn \\
& & + \,  i \ B \left[ 2 \ S + 5 \ T_1 T_2 T_3 + 2 \ (U_1
+ U_2 + U_3 ) - 3 \ ( T_1 + T_2 + T_3) \right] \, .
\label{2ads4}
\eea
Notice that condition (\ref{cond}) relates the terms with even and odd
powers of the fields in the superpotential, thus its sign ambiguity is
irrelevant. The superpotential (\ref{2ads4}) leads to a supersymmetric
vacuum at $\langle S \rangle = \langle T_A \rangle = \langle U_A
\rangle = 1$ ($A=1,2,3$). Since at this point $\langle W \rangle = 4
\, ( 3 \, A + i \, B) \ne 0$, implying $\langle V \rangle = - 3 \,
m_{3/2}^2 <0$, this vacuum has a stable $AdS_4$ geometry with all
seven main moduli frozen.

The educated reader might feel uncomfortable with condition
(\ref{cond}), which seems to imply non-integer flux numbers. This is a
consequence of our choice for presenting the model, with $S=T_A=U_A=1$
at the minimum. One can recover integer flux numbers by rescaling
appropriately the moduli. A possible choice (among many others) is the
following:
\be
(S,T_A,U_A) \; \to \; b \, (S,T_A,U_A) \, ,
\qquad 
b = \frac{B}{A} = \sqrt{3 \over 5} \, .
\ee
With that choice
\bea 
W & = & N \left[ 2\, S\left(T_1+T_2+T_3\right) -
 \left(T_1T_2 + T_2T_3 + T_3T_1\right)
 \right.  \nn \\
& & + \left.
    6\left(T_1U_1 +T_2U_2 + T_3 U_3 \right) -15 \right]  \nn \\
& & + i N \left[ 2S + 3 T_1T_2T_3 + 2 \left(U_1+U_2+U_3\right)
-3\left(T_1+T_2+T_3\right) \right] \, ,
\eea
where $N = (3/5) \, A$. 

We should emphasize here that this is the \emph{only} known example of
a complete stabilization of the moduli, reached in IIA by switching on
fundamental fluxes (NS or R). We should also stress that this cannot
happen in the heterotic string, because of the absence of
$S$-dependence in the general flux-induced superpotential. Such a
dependence could however be introduced under the assumption of gaugino
condensation. In type IIB with D3-branes, the orientifold projection
that accompanies the $Z_2 \times Z_2$ orbifold projection eliminates
the $\omega_3$ fluxes, thus the $T$ moduli are not present in the
superpotential and cannot be stabilized by fluxes. The case of
D9-branes (open string) is similar to the heterotic case, whereas the
D7-brane set-up is not captured by the $Z_2 \times Z_2$ orbifold
projection used here. The heterotic approach \`a la Horava--Witten is
under investigation \cite{dkpzlong}, whereas F-theory on Calabi--Yau
four-folds can introduce exponential dependences in the superpotential
\cite{npw} and stabilize the $T$ moduli.


\section{Conclusions and outlook}
\label{secconc}

In this paper we proposed a novel, bottom-up approach for studying the
infrared physics of superstring compactifications that preserve an
exact or spontaneously broken $N=1$ supersymmetry. The approach is in
principle applicable to all ten-dimensional superstring theories and
M--theory, and is based on the powerful constraints of the gauged
$N=4$ supergravity underlying all these compactifications. Since in
$N=4$, $D=4$ supergravity the manifold of the scalar fields is unique,
once the number of vector fields is given, our approach allows to
identify unambiguously the K\"ahler potential of the $N=1$, $D=4$
effective theory. The various systems of fluxes allowed in the
different superstring theories are then used to determine, without
solving the ten-dimensional equations of motion and Bianchi
identities, the structure constants and duality phases that specify
the gauging of the $N=4$ theory. This in turn can be used to identify
the superpotential of the resulting $N=1$ theory. The search for the
possible vacuum structures corresponding to the different systems of
fluxes can then be performed in a very powerful and elegant formalism,
by looking at the potential and auxiliary fields of the effective
$N=1$, $D=4$ theory.

To be specific, we applied our strategy to situations where the
reduction from $N=4$ to $N=1$ is achieved by a $Z_2 \times Z_2$
orbifold projection. In the present work, we kept for clarity the six
$N=4$ vector-multiplet geometrical moduli. Thus, the moduli sector of
the resulting $N=1$ theory (after the projection) contained seven
distinguished chiral multiplets: $S, T_A, U_A$, $(A=1,2,3)$. In the
heterotic theory, the $Z_2 \times Z_2$ projection is enough to reduce
the initial ($N=4$) supersymmetry to $N=1$. For describing type-II
theories, and in particular the type-IIA compactifications on which we
focused for this paper, an extra $Z_2$ orientifold projection is
needed: we chose the one acting as a parity on three of the six
internal coordinates, associated with $N=1$ compactifications of the
IIA theory with D6-branes and O6-planes. Type IIB can be treated
similarly, by introducing a $Z_2$ orientifold projection associated
with either D3-- or D9--branes.


A major geometrical difference exists, however, which makes the type
IIA compactifications far more interesting. In type IIB, the
Calabi--Yau smooth manifold resolution of the orbifold holds even in
the presence of $H_3$ and $F_3$ fluxes \cite{IIBflu}. This explains
why most of the literature deals with this kind of type IIB
constructions. However, even in this case, the $\omega_3$ geometrical
fluxes are incompatible with the $Z_2 \times Z_2$ orbifold projection
(and its smooth Calabi-Yau resolution). The situation for type IIA is
even more exotic, since no well-defined mathematical framework has yet
been unravelled for understanding the ``deformed Calabi--Yau''
geometry.

Nevertheless, this is by no means an obstruction to us, when the
situation is considered from the conformal field theory perspective of
string theory. Many examples exist that demonstrate this: asymmetric
orbifolds, fermionic constructions, twisted Gepner constructions,
supersymmetric compactifications on manifolds with torsion,
etc. Furthermore, these exact models, as well as the whole procedure
we have developed so far for dealing with fluxes, point towards the
existence of a generalized mirror symmetry, despite the absence of a
Calabi--Yau geometrical interpretation for type IIA. A manifestation
of that symmetry emerges, for instance, when a non-trivial $\omega_3$
flux is switched on. In the (freely-acting) orbifold limit, a
mirror-like $U \leftrightarrow T$ duality relates the type IIA to the
type IIB side.


The gauging approach we propose here relies on the rich but
constrained $N = 4$ structure. It enables us to bypass the above
geometrical difficulties and to organize the fluxes in a systematic
way. Indeed, the gauging procedure goes along with a set of structure
constants and duality phases, where the former must satisfy Jacobi
identities and antisymmetry conditions. These all enter the
superpotential, which in turn determines the scalar potential. In the
above framework, it is possible to list exhaustively the various
choices for the structure constants and duality phases. The choice of
the $Z_2 \times Z_2$ projection plays an important role, since it
naturally induces an interchange symmetry among the three planes, and
simplifies considerably the implementation of the full antisymmetry of
$f_{RST}$. Releasing the assumption of plane-interchange symmetry,
changing the orbifold and/or orientifold projections preserving $N=1$
supersymmetry, moving to type-IIB or type-I superstring theories, all
goes beyond the scope of the present paper, and is postponed to
\cite{dkpzlong}.

For each of the possible choices of the structure constants (this is
equivalent to choosing the subgroup of $SO(6,6)$ that is gauged) and
duality phases, one can readily analyze the issue of moduli
stabilization. Furthermore, and this is the core of the present paper,
one can trace back the origin of the structure constants and duality
phases in terms of fluxes in the underlying fundamental theory in ten
dimensions. Although our main motivation was the analysis of the yet
not unravelled type IIA, we applied our technique to the heterotic
case, where we clarified the case where geometrical and NS-NS
three-form fluxes are combined. Our pattern allows to reproduce
systematically the various examples available in the literature, such
as the no-scale models.

As far as type IIA is concerned, more possibilities exist, thanks
to the presence of R-R fluxes, besides the geometric and NS-NS
ones: $F_0$, $F_2$, $F_4$ and even $F_6$. They can be introduced
one by one, or in combination, provided the Jacobi identities are
still satisfied. A specific combination exists, which generates a
solution where \emph{all} seven moduli are stabilized, in an
$AdS_4$ geometry. This is typical of type IIA and cannot happen in
heterotic, where the allowed geometrical and three-form fluxes
cannot create an $S$-dependence in the superpotential. More
examples can be displayed with partial moduli stabilization:
domain-wall solutions, runaway solutions, no-scale models \dots

There are various directions that would be worth exploring, besides
those already mentioned above. Among them, the detailed correspondence
of the ten-dimensional equations of motion, Bianchi identities
and tadpole cancellation conditions, with the equations and
consistency conditions in the effective gauged four-dimensional
supergravity theory. Also, the inclusion in our formalism of the
scalar and vector fields associated to brane excitations, and the
exploration of the new systems of fluxes associated with their field
strengths.

\vfill{
\section*{Acknowledgments}
We thank C.~Angelantonj, C.~Bachas, M. Gra\~na, J.~Louis, R.~Minasian,
G.~Villadoro and especially S.~Ferrara for discussions. We also thank
G.~Villadoro for pointing out an incorrect sign in the original
version of eqs.~(44) and (49). J.-P.~D.  thanks the Laboratoire de
Physique Th\'eorique, Ecole Normale Sup\'erieure and the Centre de
Physique Th\'eorique, Ecole Polytechnique, for their kind hospitality
and support during the early stages of this work. C.~K., M.~P. and
F.~Z. thank the Institut de Physique, Universit\'e de Neuch\^atel for
its warm hospitality and support during parts of this
project. F.~Z. also thanks the Laboratoire de Physique Th\'eorique,
Ecole Normale Sup\'erieure for its kind hospitality and support in the
initial phase of this work. This work was supported in part by the
European Commission under the contracts HPRN-CT-2000-00131,
HPRN-CT-2000-00148, MEXT-CT-2003-509661, MRTN-CT-2004-005104 and
MRTN-CT-2004-503369.}
\end{document}